\documentclass[prb, reprint, preprintnumbers, footinbib, showpacs, citeautoscript]{revtex4-1}

\pdfoutput=1

\preprint{In Preparation}

\usepackage[T1]{fontenc}
\usepackage[english]{babel}

\usepackage{amsmath}
\usepackage{amssymb}

\usepackage{textcomp}
\usepackage{times}
\usepackage{txfonts}

\usepackage{braket}
\usepackage{graphicx}

\usepackage[colorlinks, urlcolor=blue, citecolor=blue, linkcolor=blue, pdfstartview=FitH]{hyperref}
\usepackage[all]{hypcap}

\usepackage[dvipsnames]{xcolor}

\begin{document}

\def\affiSOLAB{Spin Optics Laboratory, Saint~Petersburg State University, 198504 St.~Petersburg, Russia}
\def\affiBremen{Institut f\"ur Festk\"orperphysik, Universit\"at Bremen, 28359 Bremen, Germany}
\def\affiBochum{Angewandte Festk\"orperphysik, Ruhr-Universit\"at Bochum, 44780 Bochum, Germany}
\def\affiPaderborn{Department Physik, Universit\"at Paderborn, 33098 Paderborn, Germany}

\title{Reconstruction of nuclear quadrupole interaction in (In,Ga)As/GaAs quantum dots\\ 
observed by transmission electron microscopy}

\author{P.~S.~Sokolov}
%\altaffiliation[On leave to ]{Experimentelle Physik 2, Technische Universit\"at Dortmund, D-44221 Dortmund, Germany.}
\author{M.~Yu.~Petrov}
\email[The correspondence have to be sent to ]{m.petrov@spbu.ru}
\affiliation{\affiSOLAB}
\author{T.~Mehrtens}
\author{K.~M\"uller-Caspary}
\author{A.~Rosenauer}
\affiliation{\affiBremen}
\author{D.~Reuter}
\affiliation{\affiPaderborn}
\author{A.~D.~Wieck}
\affiliation{\affiBochum}

\date{\today}

\begin{abstract}
A microscopic study of the individual annealed (In,Ga)As/GaAs quantum dots is done by means
of high-resolution transmission electron microscopy. The Cauchy-Green strain-tensor component 
distribution and the chemical composition of the (In,Ga)As alloy are extracted from the 
microscopy images. The image processing allows for the reconstruction of the strain-induced 
electric-field gradients at the individual atomic columns extracting thereby the magnitude 
and asymmetry parameter of the nuclear quadrupole interaction. Nuclear magnetic resonance 
absorption spectra are analyzed for parallel and transverse mutual orientations 
of the electric-field gradient and a static magnetic field.
\end{abstract}


\maketitle

\section*{Introduction}

The spin physics of semiconductors has been developed for bulk materials 
and demonstrated a wide variety of linear and nonlinear phenomena, realized 
thanks to the optical orientation~\cite{OO}. It has been reborn 
in semiconductor nanostructures in recent decades~\cite{SpinPhys}.  
A significant part of the spin-related phenomena are underlain on the dynamic spin 
polarization of the nuclear spins being polarized by means of the transfer of the 
photon angular momentum to the nuclear-spin system via electron-nuclear
hyperfine interaction~\cite{LampelPRL68, DPJETP73}.
The achievement of a high polarization of the nuclear-spin system becomes 
challenging for quantum dot (QD) systems where a single spin of an electron 
would be strongly localized\cite{GammonSci96} and is under the influence 
of the nuclear spin fluctuations paving the way to a fast carrier-spin 
relaxation~\cite{GammonPRL01, MerkulovPRB02, KhaetskiiPRL02, CoishPRB04, 
JohnsonNat05, KoppensSci05, BraunPRL05}.
In spite of a combination of a large variety of methods tried to be used for 
reaching a sufficiently high degree of nuclear-spin order~\cite{LaiPRL06, 
BraunPRB06, GreilichSci07, MaletinskyPRB07, TartakovskiiPRL07, OultonPRL07, 
CarterPRL09, XuNat09, MaletinskyNatPhys09, LattaNatPhys09, VinkNatPhys09, 
CherbuninPRB09, ChekhovichPRL10, KrebsPRL10, IsslerPRL10, HogelePRL12, 
PuelbaPRB13}, the experimental achievement of the spin polarization, close 
to hundred percent, is still a challenging problem, limited, in some particular cases, 
by the quantum nature of the spin system~\cite{ImamogluPRL03, ChristPRB07, PetrovPRB09, HildmannPRB14}. 
A further microscopic analysis including a combination of experimental methods 
like nuclear magnetic resonance (NMR)~\cite{MakhoninPRB10, FlisinskiPRB10, ChekhovichNatNano12, MunschNatNano14} 
and/or various spin noise measurements~\cite{LiPRL12, Berski15, PeddibhotlaNatPhys13} 
in combination with atomistic modeling~\cite{BulutayPRB12, BulutayPRB14} 
are highly required to gain insight into such an intriguing problem.

In self-assembled QDs, $N\sim10^5$ nuclear spins interact with the 
localized-electron spin with different strengths because of a spread of the 
electron density. However, an additional spread of the interaction exists even at nuclear-spin 
level due to the crystal lattice deformation caused by built-in strain. This introduces 
a non-homogeneous nuclear quadrupole 
interaction, changing the usual nuclear-spin dynamics~\cite{DzhioevPRL07} but also 
modifying the NMR spectrum both in the single QDs~\cite{ChekhovichNatNano12} 
and in the QD ensembles~\cite{CherbuninPRB11, KuznetsovaPRB14}. The microscopic
analysis of the quadrupole interaction is a rather complex problem while any direct 
experimental measurement of its magnitude would hardly be realized in practice.
Since only scant experimental progress can be expected thereupon, an atomistic 
analysis would at least give some microscopic information within the framework 
of a chosen model~\cite{BulutayPRB14}. 

In this paper, we address the investigation of the structural properties of 
a~single QD with respect to the analysis of nuclear quadrupole interaction.
We make use of the high-angle annular dark-field (HAADF) imaging of an individual 
self-assembled QD in a scanning transmission electron microscope (STEM). 
This allows us to resolve a crystal lattice with atomic-column 
resolution that can be used to determine the shape and chemical composition of the QD. 
The investigated sample is a heterostructure grown by molecular-beam 
epitaxy on a GaAs substrate and contains 20 layers of (In,Ga)As QDs, embedded 
in a GaAs matrix. The post-growth thermal annealing of the structure 
allows for activation of in-diffusion of Ga atoms inside the QD that reduces 
the number of structure defects and, in addition, blue~shifts the ground-state 
excitonic transition~\cite{LangbeinPRB04, PetrovPRB08, NoteSample}.
The obtained STEM images were post-processed with geometric phase 
analysis (GPA), from which the Cauchy-Green strain tensor components are extracted.
In addition, to obtain the profile of the In and Ga concentrations inside the QD, 
the local chemical composition is determined by means of atomically resolved HAADF-STEM  
and energy-dispersive X-ray (EDX) spectroscopy. These data were then used to 
reconstruct the distribution of the strain-induced electric-field gradients (EFG) 
causing the nuclear quadrupole interaction. Further calculations allow the 
interpretation of the NMR transitions that can be detected in the ensemble of such QDs~\cite{FlisinskiPRB10, CherbuninPRB11, KuznetsovaPRB14}.

The paper is organized as follows. In Sec.~\ref{sec:microscopy}, 
we present the details of the microscopic study. The results of the postprocessing 
of the microscopic images are discussed. In Sec.~\ref{sec:quadrupole}, we discuss the possibility of application of microscopic methods for an analysis 
of strain-induced nuclear quadrupole interaction and quantitative modelling of the 
NMR spectra. The main results are summarized in Sec.~\ref{sec:conclusion}.

\section{\label{sec:microscopy}HAADF-STEM imaging and strain mapping}

To obtain the crystal strain and, correspondingly, the 
quadrupole interaction, high-quality microscopic imaging is required~\cite{RosenauerSprn03}.
In particular, the image should include the investigated QD and 
a region of the surrounding GaAs matrix of much larger area than the size of the QD. 
This region is used as a bulk reference area 
for further analysis. Selection of the appropriate microscopy method 
is based on the ability to  resolve the atomic columns of the crystal and the ability to extract 
the chemical composition of the QD. Such data are obtained by analysing 
the heterostructure with STEM using HAADF-detector, providing information 
about the concentration of the atoms of different species~\cite{RosenauerIOP11,Kauko13Micron,GriebIOP13,GriebUlrmicr2013}.

\begin{figure}[t]
\includegraphics[width=.8\columnwidth,clip]{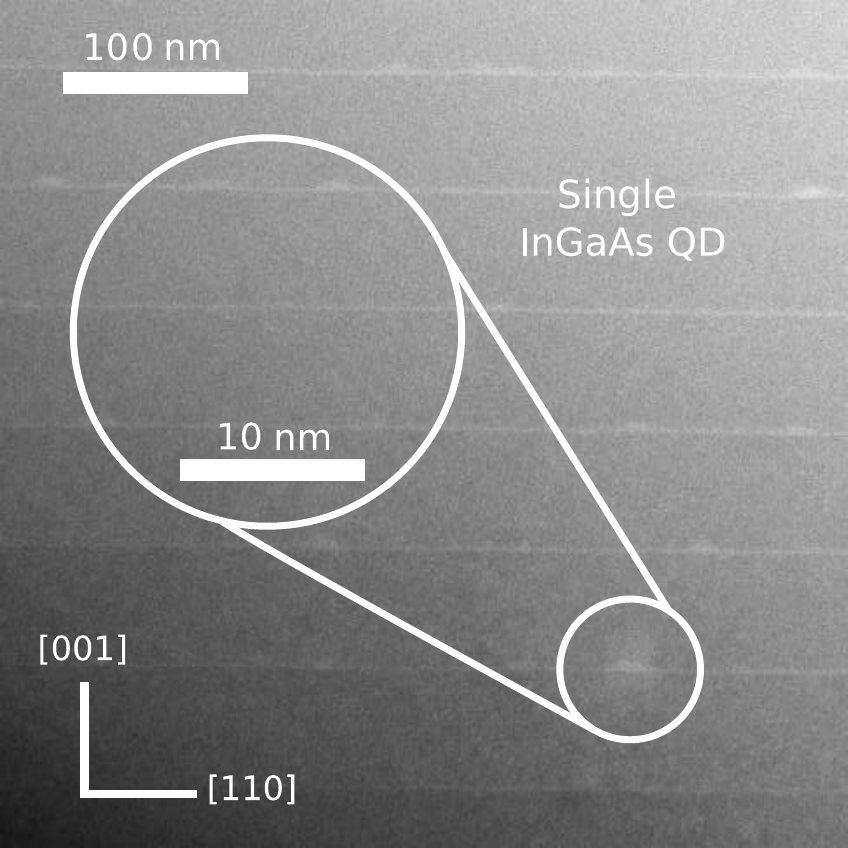}
\caption{Cross-section micrograph of the heterostructure with annealed (In,Ga)As/GaAs QDs 
obtained using scanning transmission electron microscopy. 
}\label{fig:QDLayers}
\end{figure}

The cross-sectional specimens of the sample with an area of about $0.2$~cm$^2$ 
are prepared by mechanical cutting of the heterostructure in the $(1\bar10)$ 
crystallographic plane and then glueing to a tripod holder. The specimens are then 
mechanically polished reducing the thickness of the sample down to $\sim$10--30~\textmu{}m. 
Further ion etching of the QD containing region is applied using a Gatan precision ion polishing 
system (PIPS) until a hole is etched into the sample. At the edge of this hole the sample 
thickness is clearly below 200~nm and thus transparent for electrons. The experiments are 
performed with an FEI Titan 80/300 microscope operated at 300~kV electron-beam 
acceleration voltage and equipped with an image-aberration corrector, EDX, and HAADF detectors. 
The overview image of the structure is shown in Fig.~\ref{fig:QDLayers}, from which multiple QDs 
placed on several wetting layers are resolvable. Size and shape of the dots 
are found to be weakly dispersed, which is also confirmed by a relatively narrow 
linewidth of the ground-state photoluminescence studied previously with 
these QDs~\cite{LangbeinPRB04, PetrovPRB08}. A few randomly chosen single QDs 
are analyzed.

\begin{figure}[t]
\includegraphics[width=.8\columnwidth,clip]{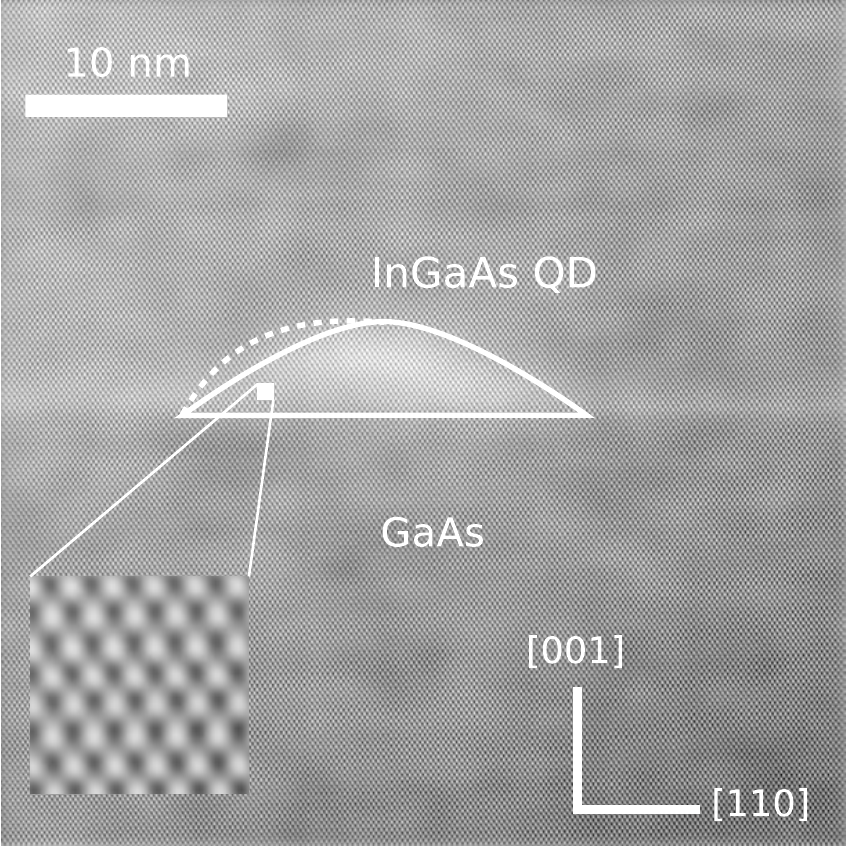}
\caption{High-resolution STEM image of a single (In,Ga)As QD embedded in the 
GaAs matrix. The shape of the QD is marked by the thin solid line guide to the eyes 
with respect to the Indium concentration profile shown in Fig.~\ref{fig:QDConcentration}. 
Atomic-column resolution is demonstrated in the inset.
}\label{fig:QDStem}
\end{figure}

Consequently, the high-resolution HAADF-STEM images were recorded in the QD-containing region 
of the sample. The image size is selected to approximately 
$50\text{\,nm}\times50\text{\,nm}$
and the spatial resolution of the microscope is around 0.12~nm in the scanning mode. Each image 
contains one QD plus a sufficient amount of the surrounding GaAs matrix, which can be used 
as reference material for thickness measurement. As high resolution was obtained in the HAADF-STEM 
images, they were not only suitable for atomic Z-contrast evaluation~\cite{GriebUlrmicr12, RosenauerUlrmicr09, MehrtensAplPhysLett13, MehrtensUlrmicr13, GrilloIOP11} 
but could be also used for geometric phase analysis~\cite{HytchUlrmicr98, CherkashinAppPhysLett13}. 
The chemically sensitive image contrast of the HAADF-STEM images (see Fig.~\ref{fig:QDStem}) 
is due to the used HAADF-detector. This ring-shaped detector detects only electrons that are 
scattered into high-angles (36--220~mrads for the camera length used in this work). The amount 
of electrons scattered into this region strongly depends on the nuclear charges of the scattering 
specimen atoms. 

\begin{figure}[t]
\includegraphics[width=.98\columnwidth,clip]{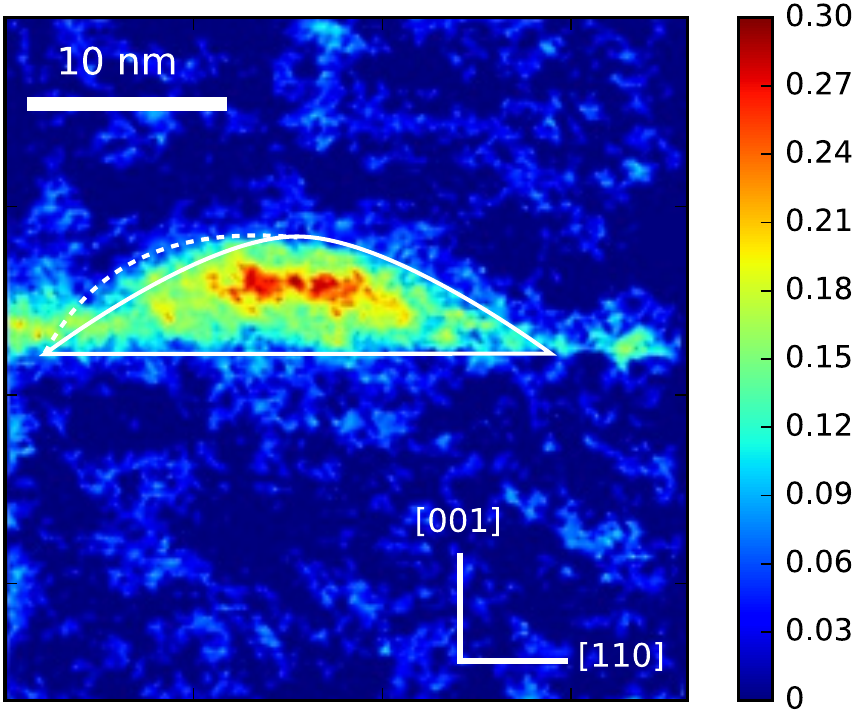}
\caption{(Color online) Indium concentration map in (In,Ga)As solid compound evaluated 
from HAADF-STEM. The solid line represents a boundary of the QD surrounding the piece of sample 
with an In-concentration higher than $0.1$ reaching the maximum value $0.3$ in the center 
of the QD. 
}\label{fig:QDConcentration}
\end{figure}

The following evaluation procedure was performed for every single image in order to obtain the 
chemical composition. First, all atomic columns in the high-resolution image were identified. 
To this end, a Wiener Filter has been applied on the original image~\cite{RosenauerSprn03, FrazhoOTAA10}. 
After that, the image has been divided into ``Voronoi''-cells and the mean intensity has been 
calculated for each ``Voronoi''-cell and has been assigned to the corresponding atomic column. 
Note that in this step the intensities of the original unfiltered image were used. To allow for the
quantitative comparison to simulated data, the mean intensity values were normalized with respect 
to the intensity of the scanning electron probe. In the following step, the normalized 
intensities were compared to reference data from multislice simulations in the frozen lattice 
approach using the STEMsim program \cite{STEMSIM} carried out in dependence of  
sample thickness and indium fraction. For more details on the simulations we refer the reader 
to Ref.~\onlinecite{MehrtensUlrmicr13}. Finally, the sample thickness was evaluated from the GaAs 
region and interpolated over the QD to allow for the determination of the indium concentration 
as shown in Fig.~\ref{fig:QDConcentration}.

To verify the accuracy of HAADF-STEM, EDX spectroscopy is used. The EDX spectra were also 
acquired in HAADF-STEM mode. While the beam is centred on a small region of the sample surface, 
the X-ray counts were integrated over one minute. Both  methods of the Z-contrast evaluation 
show very similar values for the maximum concentration of Indium in the single QD, i.\,e., 
$0.3 \pm 0.05$ and $0.35 \pm 0.05$ for EDX and HAADF-STEM, respectively.

Several methods for mapping the crystal-lattice strain can be applied~\cite{HytchUlrmicr98, CherkashinAppPhysLett13, DuggiMicrMicro14, GalindoUlrmicr07, SalesNanotech07}.
In the electron microscope, none of them is able to evaluate the strain-tensor 
component along the electron-beam propagation direction. However, the two-dimensional 
maps of two diagonal strain-tensor components could be extracted. For further 
analysis, the component of shear strain is also required. Denoting the Cartesian 
coordinate system with respect to crystallographic axes so that $x \leftrightarrow 
[110]$ and $z \leftrightarrow [001]$, the Cauchy-Green strain-tensor components 
$\varepsilon_{xx}$, $\varepsilon_{zz}$, and $\varepsilon_{xz}$ are obtained.
While other components are experimentally unaccessible, the symmetry of the problem 
allows us to equalize the $x$ and $y$ directions, keeping in mind that the specimen is 
several times thicker than the QD lateral size. We take $\varepsilon_{yy} = \varepsilon_{xx}$ 
and $\varepsilon_{yz} = \varepsilon_{xz}$ keeping $\varepsilon_{xy} = 0$. 
While $\varepsilon_{xy} = 0$ is a reasonable assumption since it is large at the heteroboundary, 
which is unsharp in the annealed QDs, other assumptions emerge as a value judgement. 
Note, that in (In,Ga)As/GaAs QDs, there is a small inequivalence of $[110]$ and $[1\bar{1}0]$
directions revealing in the exciton fine-structure splitting~\cite{LangbeinPRB04} that 
can be changed in strained QDs, as advanced atomistic modeling shows~\cite{SinghPRL10, SinghPRB13}.
%because of a small exciton fine-structure splitting arising from the inequivalence of $[110]$ and $[1\bar{1}0]$ directions~\cite{LangbeinPRB04}. 
%However, this would be taken into account by means of an advanced atomistic modeling~\cite{SinghPRL10}, 
%which is far beyond of the scope of this paper.

\begin{figure*}[t!]
\includegraphics[width=.92\textwidth,clip]{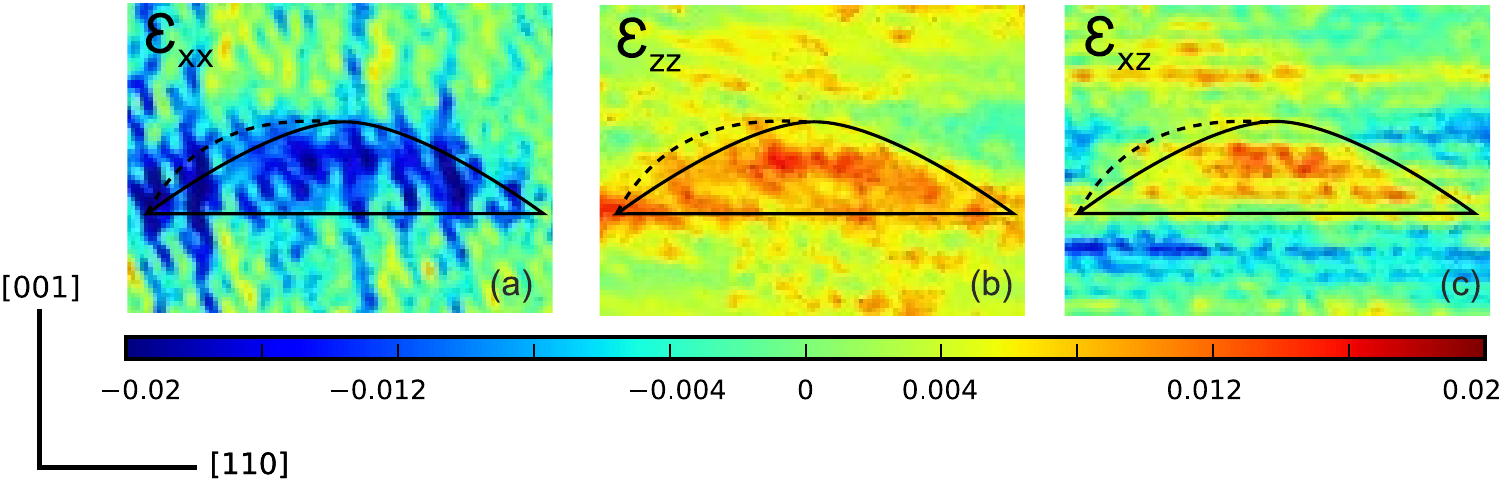}
\caption{(Color online) Cauchy-Green strain tensor components extracted from the GPA analysis of HAADF-STEM images of a single QD, from left to right: (a) $\varepsilon_{xx}$, (b) $\varepsilon_{zz}$, and (c) $\varepsilon_{xz}$.
}\label{fig:QDStrain}
\end{figure*}  

The strain mapping is performed by means of two methods, the geometric-phase 
analysis (GPA)~\cite{HytchUlrmicr98} and a modified peak-pair analysis (PPA)~\cite{GalindoUlrmicr07} 
for verification. The GPA allows us to extract the crystal-lattice distortions locally 
at each atomic column with respect to the unstrained lattice area. The basic idea 
of such an analysis is based on the fast Fourier transform of the real-space image 
into the reciprocal space. Local strain tensor components, symmetric
and rotation parts of distortion are
calculated by derivation of the displacement obtained from
two non-collinear Fourier components~\cite{GalindoUlrmicr07}. 
The phase component of this function, called the geometric phase, describes the 
position-dependent lattice deviation with respect to a reference. The reference area 
of the crystal lattice is taken in the barrier region $40$~nm away of the QD where 
the In concentration is negligibly small. The GPA strain-tensor components are 
extracted as 
\begin{equation}
\hat\varepsilon^\mathrm{GPA} = 
-\frac{1}{2\pi} \mathbf{G}^{-1} \nabla\Phi_g(\mathbf{r}),
\label{eq:GPAStrain}
\end{equation}	
where $\mathbf{G} = \left[\begin{smallmatrix} g_{1x} & g_{1z} \\ g_{2x} & g_{2z}
\end{smallmatrix}\right]$ is the matrix formed by the components of two
non-collinear reflexes $\mathbf{g}_1$ and $\mathbf{g}_2$, each of which 
is connected with the position-dependent geometric phase, $\Phi_g(\mathbf{r})$, 
as $\Phi_g(\mathbf{r}) = 2\pi \Delta \mathbf{g}(\mathbf{r}) \cdot \mathbf{r}$, 
where $\mathbf{g}(\mathbf{r})$ represents the periodicities corresponding 
to the Bragg reflections. It has the following relationship with the lattices fringe 
spacing: $d=1 / \vert \textbf{g} \vert$.

Additionally, the same STEM image with a single QD is processed using the PPA. 
In contrast to the GPA, the PPA is a real space procedure for strain mapping. 
PPA works with images having well-resolved fringe patterns, finding pairs of peaks along 
a preselected direction and distance in the affine transformed space defined by a pair 
of basis vectors $\mathbf{a}=(a_x,a_y)$ and $\mathbf{b}=(b_x,b_y)$. When the reference vectors 
are chosen on the filtered image, they can be used to construct an affine 
transformation. The next step in the PPA is the identification of peak-pairs using 
the chosen basis vectors and the intensity maxima set in the image. The Cauchy-Green 
strain components can be calculated by solving the following set of linear equations 
\begin{subequations}
\label{eq:PPAStrain}
\begin{align}
\begin{bmatrix} \varepsilon_{xx} \\ \varepsilon_{xz} \end{bmatrix}^\mathrm{PPA} &= 
{\begin{bmatrix} a_x & a_z \\ b_x & b_z \end{bmatrix}}^{-1} 
\begin{bmatrix} u_x \\ v_x \end{bmatrix},\\
\begin{bmatrix} \varepsilon_{zx} \\ \varepsilon_{zz} \end{bmatrix}^\mathrm{PPA} &= 
{\begin{bmatrix} a_x & a_z \\ b_x & b_z \end{bmatrix}}^{-1} 
\begin{bmatrix} u_z \\ v_z \end{bmatrix},
\end{align}
\end{subequations}
with coordinates of the displacements $(u_x, u_y)$ and $(v_x, v_y)$ with respect to the
reference vectors $\mathbf{a}$ and $\mathbf{b}$. 

The methods described above are based on the same general assumption that the relation 
between the atomic column positions in real crystal and features in the STEM image 
is constant inside the studied region, where the phase shift between maxima and atoms 
is supposed to be constant. Both of these two methods, within the calculation error, gave 
identical results for all three tensor components $\varepsilon_{xx}$, $\varepsilon_{zz}$, 
and $\varepsilon_{xz}$ for the single QDs under study. The PPA requires less amount of memory 
and calculation time, given that the two-dimensional complex Fourier transform is not 
required. On the other hand, PPA fails when lattice peaks are not easily detected 
due to lower resolution of the image and appearance of sublattices due to a structure defect. 
In this case, filtering of the original image can be performed. Summarizing, we 
consider that both algorithms, GPA and PPA are useful for strain mapping, each having 
different advantages and limitations, and should be considered in each particular case 
specifically. In this work, we present the results of GPA only.

In general, the strain is measured with respect to the bulk lattice parameters 
of the material. Following this definition, the GPA(PPA) strain is connected 
with the material strain, $\varepsilon_{ij}$, as follows
\begin{equation}
\Bigl(1 + \varepsilon_{ij}^\mathrm{GPA(PPA)}(\mathbf{r})\Bigr) = \Bigl(1 - \varepsilon_{ij}(\mathbf{r})\Bigr) \frac{a^\mathrm{(In,Ga)As}(\mathbf{r})}{a^\mathrm{GaAs}}.
\end{equation}
Here, the position-dependent lattice constant is determined by taking into account 
the concentration dependence of the solid compound via Vegard's law, 
$a^\mathrm{(In,Ga)As}(\mathbf{r}) = a^\mathrm{GaAs} \cdot \bigl(1-c(\mathbf{r})\bigr) + a^\mathrm{InAs}\cdot c(\mathbf{r})$, 
where $c(\mathbf{r})$ is the position-dependent In concentration shown in Fig.~\ref{fig:QDConcentration}.

Figure~\ref{fig:QDStrain} illustrates the components of the physical strain, 
$\varepsilon_{xx}$, $\varepsilon_{zz}$, and $\varepsilon_{xz}$ evaluated from GPA analysis.
The strain-tensor components extracted from high-resolution STEM images are averaged 
over the sample thickness, i.e., their values in the QD would be slightly larger in 
magnitude. Note, that an additional source of strain release arises from the sample thinning 
down to less than a \textmu{}m. In this case, the QD becomes closer to a surface than 
a deep-in-bulk QD. Therefore, both these effects should be considered as a source 
of measurement error for further evaluation of the results.

The negative value of $\varepsilon_{xx}$ shows that the crystal lattice 
is compressed in $x$ direction ($[110]$) [see fig. \ref{fig:QDStrain} (a)]. On the contrary, the $\varepsilon_{zz}$ 
component is mostly positive in the upper left region of the QD (see fig. \ref{fig:QDStrain} (b)), i.e., there is 
a stretching of the lattice in the growth direction $z$ ($[001]$). 
Fig.~\ref{fig:QDStrain}(c) shows a shear-strain $\varepsilon_{xz}$, which is 
present in a relatively small area having a mostly positive value and some 
periodicity in its distribution. In the wetting layer, the values of shear strain are 
predominantly negative, while inside the QDs this deformation is positive. Overall, 
the obtained results qualitatively well coincide with the finite-element modelling, represented 
previously in Ref.~\onlinecite{KuznetsovaPRB14}.

\begin{figure}[b]
\includegraphics[width=.98\columnwidth,clip]{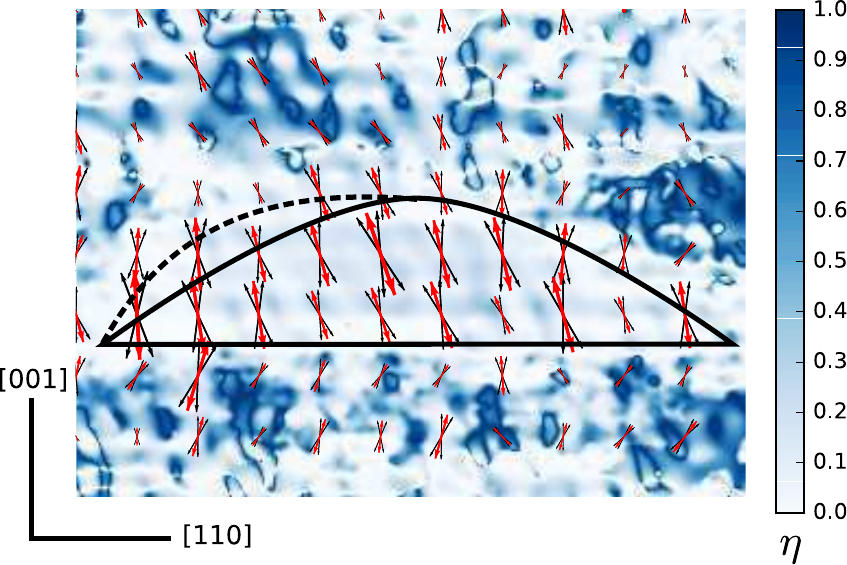}
\caption{(Color online) EFG averaged over the propagation direction of the electron beam. The red arrows indicate the relative value of the strain-induced quadrupole frequency. All the microscopy data errors are indicated by the additional black arrows. The surface plot shows the spatial distribution of the EFG asymmetry $\eta$. Black lines indicate the shape of the QD extracted from the (In,Ga)As alloy concentration map.
}\label{fig:QDQuadrupole}
\end{figure}

\begin{figure*}[t]
\includegraphics[width=.44\textwidth,clip]{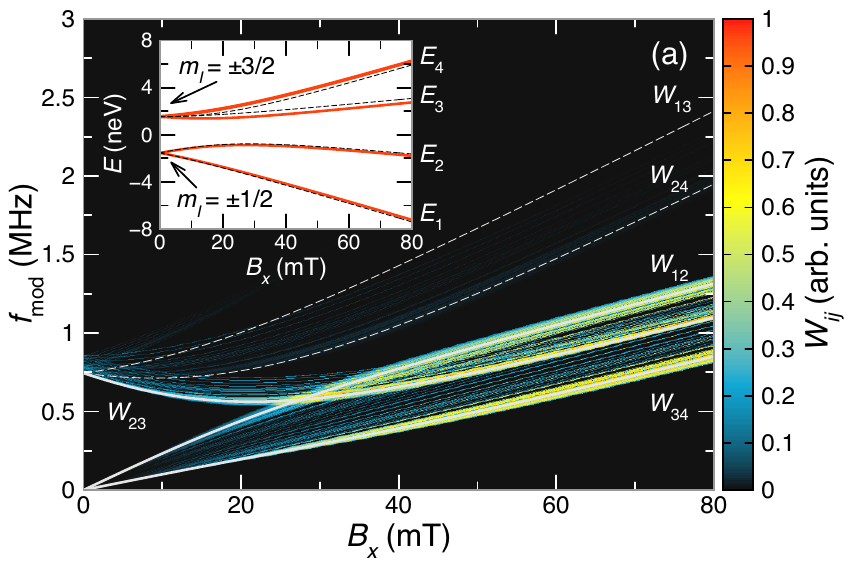}
\includegraphics[width=.44\textwidth,clip]{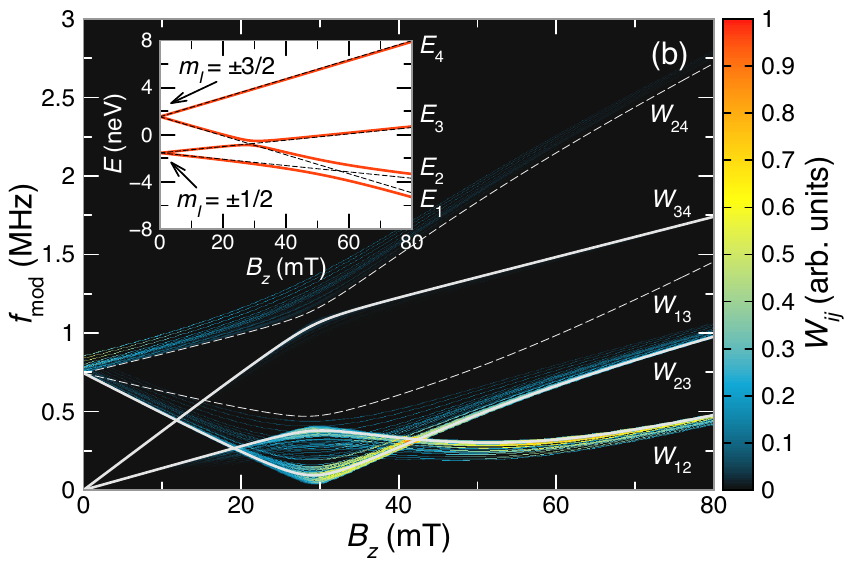}
\caption{(Color online) The field-frequency maps of the nuclear spin-flip transition rates given by Eq.~\eqref{eqRates} calculated when the magnetic field is oriented perpendicular (a) or parallel (b) to the QD growth axis. The raster map indicates the probability of the transitions for $^{71}$Ga nuclei with spin $I=3/2$ (as an example). The transitions between dipole-allowed and dipole forbidden states calculated with $\eta$ and $V_\mathcal{ZZ}$ over the QD volume are marked by solid and dashed lines, respectively. The inset graphs demonstrate the splittings of the nuclear-spin states for a single nucleus affected by a mean $\nu_Q$ in symmetric (dashed lines) or asymmetric (solid lines) quadrupole configurations.
}\label{fig:ODNMR}
\end{figure*}

\section{\label{sec:quadrupole}Analysis of strain-induced quadrupole interaction}

The change of energy experienced by the nuclear spin $\hat{\mathbf I}$ can be dependent 
of nuclear orientation. The charge environment of the nuclei interacts with the external electric
potential $V$. In the equilibrium, the nuclei experience zero average electric field but nuclei 
having $I>1/2$ have also nonzero quadrupole moment interacting with the EFG denoted hereafter as
$V_{ij}$~\cite{AbragamBook, SlichterBook}. The physical origin of nonzero $V_{ij}$ is any 
inhomogeneity of the electric fields. The strongest effect is caused by the substitutional atoms 
in the alloy. Due to different electronegativity of the cation atoms, the anion As nuclei are 
highly affected by the EFG caused by the charge environment redistribution. For example, 5\% 
electronegativity difference between the group-III atoms leads to the nuclear-spin splitting of 
several tens of MHz, as shown in (Al,Ga)As bulk~\cite{OO}. 

In self-assembled QDs, a weaker splitting would be obtained due to the crystal-lattice biaxial 
strain of several percent, leading to the quadrupole splitting of nuclear-spin states of the 
order of a MHz~\cite{DzhioevPRL07}. Even in the absence of both effects, a strong electron 
localization would lead to the additional inhomogeneity of spin splitting caused by interaction 
with the applied electric field $V_{ij}$~\cite{BrunPR63}. 
Our estimation shows, however, that the EFG variation due to inhomogeneity of the ground-state 
electron density inside the model QD leads to, at least, an order of magnitude smaller variation 
of spin splitting than its strain-induced value. However, in electrically driven QDs or
in self-assembled QDs with a bias-controlled charge state, the electric-field induced EFG 
variation requires additional verification, as shown previously in a quantum well system~\cite{EickhoffPRB03}.

Considering further the strain-induced quadrupole interaction, the Hamiltonian reads as
\begin{equation}
	\hat{\mathcal H} = - \hslash \gamma_I \hat{\mathbf I}\cdot\mathbf{B} + \frac{h\nu_Q}{2}\Bigl(3\hat{I}_{\mathcal Z}^2 - \hat{\mathbf I}^2 + \frac{\eta}{2}\bigl(\hat{I}_{\mathcal X}^2+\hat{I}_{\mathcal Y}^2\bigr)\Bigr).
	\label{eqQZHam}
\end{equation}
Here, the first term couples nuclear spin $\hat{\mathbf{I}}$ of the $I$th nucleus having 
the gyromagnetic ratio $\hslash\gamma_I$ with the external magnetic field 
$\mathbf{B}$. The second term describes the nuclear quadrupole interaction with 
strength $h\nu_Q$. The coordinate system $\mathcal{(X,Y,Z)}$ corresponds to the major 
frame of the EFG acting on the nucleus generally not coinciding with the $(x,y,z)$ 
frame of the crystal. 
The quadrupole frequency $\nu_Q$ is proportional to the major EFG component 
$V_\mathcal{ZZ}$ satisfying $V_\mathcal{ZZ}>V_\mathcal{YY}>V_\mathcal{XX}$. 
The quadrupole asymmetry parameter, $\eta = (V_\mathcal{XX}-V_\mathcal{YY})/ 
V_\mathcal{ZZ}$, is determined by the relative ratio of its other principal 
components.

The microscopic analysis allows to quantitatively evaluate the EFG tensor and, 
as a consequence, to map the quadrupole interaction over the sketch of the QD. 
The EFG tensor relates to the elastic strain as $V_{ij} = S_{ijkm}\varepsilon_{km}$, where 
$S_{ijkm}$ is the fourth-rank gradient elastic tensor~\cite{SundforsPRB74}. 
Using the strain-tensor components extracted from HAADF-STEM analysis 
(see Fig.~\ref{fig:QDStrain}), we calculate all non-zero components of $V_{ij}$, 
and, correspondingly 
\begin{equation}
h\nu_Q = \frac{e Q V_\mathcal{ZZ}}{4I(2I-1)}	
\end{equation}
where $Q$ is the nuclear quadrupole moment.
The results of such an analysis for $^{71}$Ga are plotted in Fig.~\ref{fig:QDQuadrupole}. 
The magnitude of $\nu_Q$ varies by the order of magnitude over the QD volume. 
However, the principal direction $\mathcal{Z}$ of the EFG varies by approximately 
10 degrees, keeping a mean value misalignment of 8 degrees from the $[001]$ 
crystal axis (see Fig.~\ref{fig:QDQuadrupole}). The variation of the EFG asymmetry 
$\eta$ is found to be less than 0.2 over the whole QD volume.

To quantitatively evaluate the NMR absorption, the eigenvalue decomposition 
of Eq.~\eqref{eqQZHam}, taking into account the spatial variation of $\nu_Q$ 
and $\eta$, is made. The spin-flip rates between $|i\rangle$ and $|j\rangle$ 
eigenstates having energies $E_i$ and $E_j$ are expressed as follows~\cite{SlichterBook} 
\begin{equation}
W_{ij} = \frac{2\pi}{\hslash}\left|\langle i| \hat{I}_z| j\rangle\right|^2 
\delta(E_i-E_j-hf_\mathrm{mod})
\label{eqRates}
\end{equation}
when the $z$~axis coincides with the quantization axis of electron-spin observable.
The transitions can be induced by either the absorption of radio-frequency 
electromagnetic fields or by the temporal variation of the Knight field due to 
electron-spin pumping, both modulated at frequency $f_\mathrm{mod}$. The eigenstates 
$|i\rangle$ and $|j\rangle$ are no more eigenstates of $\hat{I}_z$, therefore 
the transitions with momentum projection by $\Delta m_I>1$ are allowed under certain 
conditions~\cite{EickhoffPRB02}. The transition rates are calculated when scanning 
the magnitude of the external magnetic field oriented perpendicular 
[Fig.~\ref{fig:ODNMR}(a)] or parallel [Fig.~\ref{fig:ODNMR}(b)] 
to the $[001]$ crystallographic axis, respectively. The statistics caused 
by the inhomogeneous variation of $\nu_Q$ and $\eta$ 
shown in Fig.~\ref{fig:QDQuadrupole} over the QD is taken into account here.

In both orientations of the magnetic field, the transition lines are strongly 
broadened due to inhomogeneity of the quadrupole interaction within the QD volume. 
A geometry with $\mathbf B$ aligned across the growth axis is more sensitive to the 
value of $\nu_Q$. Here, the Kramers doublets with $m_I=\pm 1/2$, $\pm 3/2$, etc., 
instantaneously mix in the $\mathbf{B}$ field, thus splitting the states linearly 
for $m_I=\pm 1/2$ and nonlinearly for higher states. The additional asymmetry 
of the quadrupole configuration given with $\eta>0$ or a small tilt of the magnetic 
field from exactly perpendicular to the EFG axis results in a change of the splitting, 
particularly for states having $|m_I|>1/2$. Both, the variation of $\nu_Q$ and $\eta$, 
and the magnetic-field tilting result in a broadening of the transition lines, 
as Fig.~\ref{fig:ODNMR}(a) shows. As a characteristic scale, the strongest 
transition between $m_I = \pm 1/2$ of $^{71}$Ga spreads over about several hundred kHz 
in moderate fields $B>40$~mT [see Fig.~\ref{fig:ODNMR}(a)]. If the magnetic field 
is oriented along the EFG axis, the nuclear-spin states are eigenstates 
of the Hamiltonian and do not mix by the magnetic field. However, a small tilt 
of the field results in a mixing, causing an anticrossing of lowest-energy 
$E_1$ and $E_2$ states [see inset in Fig.~\ref{fig:ODNMR}(b)]. Further mixing 
of states is provided by the asymmetric quadrupole configuration with $\eta>0$. 
Both effects lead to broadening of the sharp lines to values, however, smaller than 
in the perpendicular geometry by a factor of two, at least. This goes well with 
the recent observation of NMR lines in a single QD reported in Ref.~\onlinecite{BulutayPRB14} 
where the tail of the line of several hundred kHz is observed for all nuclear isotopes. 

The shear strain expressed in the dot as $\varepsilon_{xz}$ and providing nonzero 
$\eta$ further modifies qualitatively the NMR spectra. For the sake of comparison, 
the energy-level schemes in both symmetric and asymmetric quadrupole configurations 
are shown in the insets of Figs.~\ref{fig:ODNMR}(a) and \ref{fig:ODNMR}(b). 
The frequency shifts of several tens of kHz are observed in both geometries. 
In addition to that, the spin-level $E_2$ and $E_3$ anticrossing exists when $\eta>0$ 
[see Fig.~\ref{fig:ODNMR}(b)]. The sample tilting is a natural way to reduce 
the broadening~\cite{SlichterBook}, that works well in single-QD NMR~\cite{BulutayPRB14}. 
However, full compensation of the spin-level anticrossings is prohibited 
if $\eta \neq 0$, as demonstrated in Fig.~\ref{fig:ODNMR}(b).

\section{\label{sec:conclusion}Conclusion}

To conclude, the microscopic study of the quadrupole interaction 
in self-assembled (In,Ga)As QD is done by using scanning transmission electron 
microscopy. The HAADF-STEM technique allows us to extract the (In,Ga)As alloy concentration 
of In and Ga atoms, keeping the atomic-column resolution of the microscopy image, and 
to evaluate the in-plane components of the Cauchy-Green strain tensor 
by using the geometric phase analysis. Further mapping of the biaxial and shear 
strain components allows for quantitative reconstruction of the strain-induced 
EFG tensor components. Modelling the NMR absorption spectra, the magnitudes of 
the NMR lines broadenings and a shift of certain NMR transitions caused 
by the asymmetry of the EFG tensor are evaluated. In particular, the asymmetry parameter
of the quadrupole interaction leads to shifts and anticrossings of certain transitions
in the NMR absorption spectrum.

\begin{acknowledgements}
We would like to thank I.V. Ignatiev (Saint-Petersburg State University, Russia), A. Greilich, 
D.R. Yakovlev, and M. Bayer (Technical University of Dortmund, Germany) for the discussion 
and valuable comments on the paper.
This work is partially supported by the Russian Foundation for Basic Research 
and the Deutsche Forschungsgemeinschaft in the frame of the International Collaborative 
Research Center TRR 160 and by the EU~FET-program SPANGL4Q. 
P.S.S. would like to acknowledge support of SPbSU research grant No.~11.38.213.2014 
and the BISIP Santander Scholarships programme (Bremen International Student 
Internship) whereby the microscopic experiments were conducted at the University of 
Bremen. 
M.Y.P. acknowledges support of the Russian Ministry of Education and Science under grant
No.~11.G34.31.0067. 
A.D.W. acknowledges gratefully support of Mercur Pr-2013-0001, BMBF - Q.com-H  
16KIS0109, and the DFH/UFA  CDFA-05-06. 
K. M.-C. acknowledges support from the Deutsche Forschungsgemeinschaft (DFG) under contract No. MU 3660/1-1.
\end{acknowledgements}


\begin{thebibliography}{}
\bibitem{NoteSample} The analyzing sample was exposed to a rapid thermal annealing of 30~s 
duration at temperature $T_a=920$~\textdegree{}C. As a result of the bandgap 
renormalization, the ground state photoluminescence changes from 1.06~eV to 1.37~eV.
\end{thebibliography}
\end{document}